\documentstyle[aps,multicol,epsf]{revtex}
\newcommand{\be}{\begin{equation}}
\newcommand{\ee}{\end{equation}}
\begin{document}
\draft
\title{Spectra and eigenvectors of scale-free networks \\}
\author{K.-I.~Goh, B.~Kahng, and D.~Kim \\}
\address{
Center for Theoretical Physics and School of Physics, 
Seoul National University, Seoul 151-742, Korea \\}
\date{\today}
\maketitle
\thispagestyle{empty}
\begin{abstract}
We study the spectra and eigenvectors of the adjacency matrices 
of scale-free networks when bi-directional interaction is allowed, 
so that the adjacency matrix is real and symmetric. 
The spectral density shows 
an exponential decay around the center, followed by power-law long tails 
at both spectrum edges. The largest eigenvalue $\lambda_1$ depends 
on system size $N$ as $\lambda_1 \sim N^{1/4}$ for large $N$, and 
the corresponding eigenfunction is strongly localized at the hub, the 
vertex with largest degree. The component of the normalized eigenfunction 
at the hub is of order unity. We also find that the mass gap scales 
as $N^{-0.68}$.\\ 
\end{abstract}
\begin{multicols}{2}
\narrowtext
Complex systems such as social, biological, and economic systems 
consist of many constituents such as individuals, substrates, and 
companies, respectively\cite{complex}. Each constituent reacts 
and adapts to the pattern created in the system through diverse 
interactions.  
Cooperative phenomena between constituents in such complex systems 
may be described in terms of random graphs, consisting of 
vertices and edges, where vertices (edges) represent constituents 
(their interactions). The study of complex systems in terms of 
random graph was initiated by Erd\"os and R\'enyi (ER) \cite{er}. 
In the ER model, the number of vertices is fixed, 
and edges connecting from one vertex to another occur 
with probability $p$. Then there exists a probability 
threshold $p_c$, above which the system is percolated.    
Recently Watts and Strogatz (WS) introduced the small-world network 
\cite{ws} to consider local clustering, while the number of 
vertices is also fixed. The WS model offers the first indication 
that real networks could be more complex than predicted 
by the ER model. \\ 

Recently Barab\'asi and Albert (BA) \cite{ba} introduced an 
evolving model where the number of vertices increases linearly with time 
rather than fixed, and a newly introduced vertex is connected 
to $m$ already existing vertices, following the so-called
preferential attachment rule that the vertices with more edges 
are preferentially selected for the connection to the new vertex. 
The number of edges $k$ incident upon a vertex is the degree of the 
vertex. Then the degree distribution $P(k)$ of vertices, 
equivalent to the connectivity distribution, follows a power-law 
$P(k) \sim k^{-3}$ for the BA model, while for the ER 
and WS models, it follows a Poisson distribution. 
The BA model is interesting in a sense that a lot of complex 
interactive networks such as the world wide web \cite{www1,www2}, 
the actor network \cite{ba}, the citation network of scientific 
papers and the author collaboration network\cite{citation}, 
metabolic network \cite{metabolic} and food web \cite{foodweb} 
in biological systems all exhibit the power-law in the degree 
distribution, implying that a characteristic degree is absent 
in such systems. Thus the BA model is called a scale-free network. 
In the BA network, one may assign a direction on each edge pointing 
from the younger vertex to the older one\cite{porto}. However, when 
the direction on each edge is ignored, allowing bi-directional 
interactions such as download and upload communications in Internet, 
the BA network may be regarded as a simple model for the Internet 
topology\cite{internet}.\\ 

While it is known that the BA network follows the power-law 
in its degree distribution, further structural properties are 
not so well known. When $m=1$, the BA 
network forms a tree structure without forming any loop, but 
for $m > 1$, loops are formed, and network topology becomes much 
complicated. So it would be interesting to investigate the spectrum 
of the BA network, because generally the spectrum of a random 
graph and corresponding eigenvectors are closely related to 
topological features of the random graph \cite{book1,book2,book3}. 
In this paper, we study the spectrum and the corresponding 
eigenvectors of the adjacency matrix of the BA network, 
comparing spectral properties with structural features. 
When a BA network is composed of $N$ vertices, the adjacency 
matrix $\bf A$ consists of $N \times N$ elements, 
$\{a_{i,j}\}$ ($i,j=1,\dots,N$), defined as $a_{i,j}=1$ if 
vertices $i$ and $j$ are connected via an edge, and $a_{i,j}=0$ 
otherwise. $k_i=\sum_j a_{i,j}$ is the degree of the vertex $i$. 
In the BA model, the vertex with the largest degree is singled out 
and is called the hub, denoted by $h$ hereafter, in this work. 
From previous studies~\cite{ba}, we know that $k_h$, the degree 
of the hub, scales as $k_h \sim mN^{1/2}$.
The vertices are ordered in their ages; the vertex $i$ is the one 
born at time $i$.\\ 

Since we consider the bi-directional case (sometimes called the undirected 
case), the adjacency matrix $\bf A$ is real and symmetric, so that 
all eigenvalues are real and the largest eigenvalue is not degenerate. 
We obtained the spectrum of the BA network via exact diagonalization 
for $N$ up to $5,000$ and for the first few largest eigenvalues, via the 
Lanczos method \cite{lanczos} for $N$ up to $400,000$. 
Throughout this paper, numerical simulations were carried out 
for $m=2$, the simplest case including the loop structure.\\

$Eigenvalues-$ We consider the distribution of the eigenvalues, 
shown in Fig.1. Analytic formula for the spectrum is not known yet, but one 
can easily see that the spectrum does not fit to the 
semi-circular equation derived by Wigner\cite{book1} appropriate to the ER 
random graph. The density of eigenvalues $\rho(\lambda)$ 
in the middle part of the spectrum is likely to fit to the 
formula, $\rho(\lambda)\sim \exp(-|\lambda|/a)$ where 
$a\approx 1.25$ (see the left inset of Fig.1), while the density 
further out follows the power-law, $\rho(\lambda)\sim |\lambda|^{-4}$, 
(see the right inset of Fig.1). Since the power-law decays much slower 
than the exponential one, the spectrum shows long tails at both 
spectrum edges (see Fig.1). Such behavior has also been observed 
in the financial time series \cite{matrix}. \\

The size dependence of the largest eigenvalue $\lambda_1$ is of 
interest and we found numerically that $\lambda_1$ increases as 
$\sim N^{1/4}$ for large systems (see Fig.2). 
To consider the relation between the largest eigenvalue 
$\lambda_1$ and the structure of the BA 
network, we consider the followings. 
Let $\bf x$ and $\bf y$ be $N\times 1$ column vectors, related through 
\begin{equation}
{\bf y}(n)={\bf A}^n {\bf x}.
\end{equation}
Let $v_{i,l}$ be the $i$-th component of the normalized eigenvector 
corresponding to the $l$-th eigenvalue of $\bf A$, $\lambda_{l}$, 
($\lambda_1 > \lambda_2 \ge \dots \ge \lambda_N$). 
Then Eq.(1) may be rewritten as 
\begin{equation}
y_{i}(n)=\sum_{l}\sum_j \lambda_{l}^n v_{i,l} v_{j,l} x_j,
\end{equation}
When we set $x_j=1$ for some $j$ and $x_r = 0$ for $r\neq j$, 
$y_{i}(n)$ becomes the number of possible ways of $n$-step walks 
starting from the vertex $j$ and terminating at the 
vertex $i$. This will be denoted as $y_{j \rightarrow i}(n)$ hereafter. 
In particular, when $i=j$, $y_{j\rightarrow j}(n)$ is the number 
of possible ways to return to the starting vertex $j$ 
after $n$-step walks.  When $n$ is sufficiently large, 
$y_{j\rightarrow i}(n)$ can be represented in terms of the largest 
eigenvalue $\lambda_1$ alone as 
\begin{equation}
y_{j\rightarrow i}(n)\approx \lambda_1^n v_{i,1}v_{j,1}.
\end{equation}
In particular, when $i=j$, 
\begin{equation}
y_{j\rightarrow j}(n)\approx \lambda_1^n v_{j,1}^2.
\end{equation}
\\

Even though the relation Eq.(4) holds for any vertex $j$, we consider 
it particularly at the hub. 
Now let us consider the number of ways of $n+2$ steps returning 
to the starting vertex $h$. 
It can be split into two parts,  
\begin{equation}
y_{h \rightarrow h}(n+2) = y_{h \rightarrow h}(n)y_{h \rightarrow h}(2)
+\sum_{j\ne h} y_{h \rightarrow j}(n) y_{j \rightarrow h}(2). 
\end{equation} 
According to Eq.(4), 
\begin{equation}
\lambda_1^2 \approx {y_{h \rightarrow h}(n+2)\over y_{h \rightarrow h}(n)}
=y_{h \rightarrow h}(2)+{{\sum_{j\ne h} y_{h \rightarrow j}(n)
y_{j \rightarrow h}(2)}\over {y_{h \rightarrow h}(n)}}, 
\end{equation}
where $y_{h \rightarrow h}(2)$ corresponds to the degree of the hub, 
$y_{h \rightarrow h}(2)=m\sqrt{N}$, while $y_{j \rightarrow h}(2)$ 
is found numerically to behave as $\sim N^{0.05}/j^{0.43}$, weakly 
depending on $N$ (see Fig.3). The second term of the right hand side 
of Eq.(6) can be written using Eqs.(3) and (4) as 
$\sum_{j\ne h} \Big({{v_{j,1}}\over {v_{h,1}}}\Big) y_{j \rightarrow h}(2)$.
We will show later that $v_{h,1}\approx 1/2$, while   
$v_{j,1}\sim (Nj)^{-1/4}$ for $j$ being away from $h$, leading to the 
result that the second term becomes ${\cal O}(N^{0.1})$.
Consequently, $\lambda_1$ is contributed dominantly by the first term
in the right hand side of Eq.(6), leading to $\sim m^{1/2}N^{1/4}$.\\

The result of $\lambda_1 \sim N^{1/4}$ can also be understood 
through two toy models. 
First, one vertex is located at the center, and all the other $N-1$ 
vertices are linked only to the center vertex (Fig.4a). 
This structure, called the radial structure, is an extreme case 
of ``winner-takes-all". The largest eigenvalue of this structure is 
$\lambda_1=\sqrt{N-1}$, while the largest degree is $N-1$. Second, 
we consider a two-level Cayley tree structure 
(Fig.4b). When the coordination number is chosen as $k=\sqrt{N}$, 
the total number of vertices becomes $N+1$, and the largest degree 
$\sqrt{N}$, as in the case of the BA network. 
The largest eigenvalue for this toy model is found to be 
$\lambda_1=\sqrt{2k-1}\sim N^{1/4}$. So, in both models, 
$\lambda_1 \sim \sqrt{k_h}$.\\

The second largest eigenvalue in its absolute magnitude is 
located on the negative side of the spectrum. The absolute 
magnitude of this eigenvalue $|\lambda_N|$ is also 
found to scale as $|\lambda_N|\sim N^{1/4}$ (see Fig.2). 
The difference between the 
first two largest eigenvalues in their absolute magnitude, 
$\lambda_1 - |\lambda_N|$, is found to scale as $\sim N^{-0.43}$ 
(see the inset of Fig.2). 
Thus, the mass gap, defined as $\log (\lambda_1/|\lambda_N|)$ 
\cite{massgap}, scales as $\sim N^{-z}$ with $z \approx 0.68$.\\

Next, we consider the size dependence of the moments of the 
spectrum. First, since the matrix $\bf A$ is 
traceless, the first moment becomes zero, that is, 
${\cal M}_1=\sum_i^N \lambda_i=0$.
Second, for each vertex, the number of ways to return to the  
starting vertex by a 2-step walk is the same as the degree 
of that vertex, so that ${\cal M}_2=\sum_i^N \lambda_i^2=2mN$. 
Third, since the path to return to the starting vertex by a 3-step 
walk forms a triangle, the total number of triangles ${\cal T}_N$ in the 
system can be obtained through the third moment, 
${\cal M}_3=\sum_i^N \lambda_i^3=6 {\cal T}_N$. 
We found numerically that ${\cal M}_3 \sim N^{0.40}$ (see Fig.5). 
Note that while the first moment vanishes, the third moment 
does not, implying that the spectrum is not completely symmetric 
with respect to $\lambda=0$.\\

$Eigenvector-$ Since the adjacency matrix $\bf A$ is real and symmetric, 
every component of the eigenvector corresponding to the 
largest eigenvalue is positive. Since the largest eigenvalue 
and the corresponding eigenfunction is important, as in the case of 
Eq.(3), we focus on the eigenvector $\{v_{j,1}\}$ for the largest 
eigenvalue. We consider the square of each component of the eigenvector 
$\{v_{j,1}^2\}$ instead of $\{v_{j,1}\}$ itself because of the 
normalization, $\sum_j v_{j,1}^2=1$, and compare it with the 
normalized degree of each vertex $\{k_j/\sum_l k_l\}$. 
As seen in Fig.6, the two quantities are correlated in such a way 
that up-down behavior occurs in a similar fashion. \\ 

The components $\{v_{j,1}^2\}$ ($j=1,\dots,N$) is strongly localized 
at the hub. This is in contrast to the ER case where the corresponding 
eigenfunction is extended over all vertices (see the inset of Fig.6). 
We found that $v_{h,1}^2$ at the hub increases with increasing 
$N$ for small $N$, but converges to a constant close to $1/2$ for 
large $N$ (see Fig.7). The value $1/2$ can be obtained from the 
radial structure analytically. 
Since the radial structure is the extreme case that the connectivity 
is mostly localized at the hub, the value $1/2$ can become the upper 
bound of $v_{h,1}^2$. 
The result that $v_{h,1}^2$ approaches a constant as $N \rightarrow 
\infty$ is interesting, because $v_{h,1}^2$ behaves differently from 
the normalized degree at the hub, decreasing as $\sim N^{-1/2}$ 
(see Fig.7). 
Since the number of returning $n$-step walks from a vertex $j$ is 
$y_{j \rightarrow j}(n) \propto v_{j,1}^2$ from Eq.(4), $v_{j,1}^2$ 
can be considered as a measure of contribution from the vertex 
$j$ to transport processes in a network. The fact that 
$v_{h,1}^2 \sim {\cal O}(1)$ can be 
interpreted as that the hub plays much more dominant role in 
transports on the BA network, compared with the contribution measured 
by the normalized degree at the hub, which is ${\cal O}(1/\sqrt{N})$. 
Thus, the presence of the hub in the scale-free networks diversifies 
pathways, and enhances the efficiency of transport, while the network 
systems are vulnerable to the attack on the hub \cite{attack1}.  
On the other hand, for other components of the eigenfunction, 
$\{v_{j,1}^2\}$, ($j=1,\dots,N$, but $j\ne h$), 
it was found numerically that $v_{j,1}^2 \sim 1/4\sqrt{Nj}$ 
(see Fig.8).\\ 
 
Each component of eigenvectors can be obtained through the 
diagonalization of the adjacency matrix $\bf A$. 
However, the eigenfunction for $\lambda_1$ can be obtained 
using the relation, Eq.(4). The knowledge of $y_{j\rightarrow j}(n)$ 
and $\lambda_1$ enables one to obtain each component of the 
eigenfunction through $v_{j,1}^2 \approx y_{j\rightarrow j}(n)/\lambda_1^n$ 
for sufficiently large $n$. However, when $n$ is not large enough, 
$y_{j\rightarrow j}(n)$ for vertex $j$ is affected by local topology 
around $j$, so that the value of $v_{j,1}^2$ obtained by this method 
would be different from that by the exact diagonalization. 
Thus, we can define a characteristic number of steps $n_c$, 
such that for $n > n_c$, Eq.(4) holds, while for $n < n_c$, it 
breaks down. To find $n_c$, we introduce the quantity, 
\begin{equation}
\delta_n\equiv \Big | \sum_j 
\Big({y_{j\rightarrow j}(n)\over \lambda_1^n}-v_{j,1}^2 \Big)
\Big |,
\end{equation}
which is found to decay exponentially as $\delta_n\sim \exp (-n/n_c)$ 
(see the inset of Fig.9).
We found numerically that $n_c$ shows a crossover across a characteristic 
system size $N_c$ such that for $N < N_c$, $n_c \sim N^{0.35}$, while for 
$N > N_c$, $n_c \sim N^{0.50}$ (see Fig.9). On the other hand, Eq.(7) may 
be rewritten as $\delta_n =|\sum_{l >1} ({\lambda_{l} / \lambda_1})^n|.$
Then, for sufficiently large $n$, $\delta_n$ is contributed dominantly 
by the term $(|\lambda_N|/\lambda_1)^n$ alone. Combined with 
$\delta_n \sim \exp(-n/n_c)$, one obtains that 
$n_c = \lambda_1/(\lambda_1-|\lambda_N|)$, equivalent 
to the inverse of the mass gap. Thus, $n_c \sim N^{z}$. 
However, the numerical value of the exponent $z$ 
obtained in this way, $z\approx0.50$, deviates from the value 
$z \approx 0.68$ obtained by the direct measurement of 
$\lambda_1-|\lambda_N|$. This can be attributed to the fact that $\delta_n$ 
includes contributions from other eigenvalues. \\

$Shortest$ $paths-$ Since the role of the hub in transport is 
much more dominant, compared with the contribution by the normalized degree, 
we also study topological feature of shortest paths between two vertices
\cite{short}.  
The transport from one position to another is mainly carried along 
the shortest path(s) between them, and is contributed dominantly by the 
largest eigenvalue of the adjacency matrix $\bf A$. 
We define a set, composed of the vertices on the shortest path(s) 
from one position to another. Then, since there are $N(N-1)/2$ 
pairs of vertices, the same number of sets exist in the system. 
Among them, we are interested in how many number of different sets 
a certain vertex $j$ belongs to. This number is called the involving 
number $P_j$, while the normalized involving number is defined as 
$p_j \equiv P_j/\sum_{l} P_{l}$ for each vertex $j$. Fig.6 also shows 
$p_j$ versus $j$. It behaves similarly to the up-down behavior of
$v_{j,1}^2$ and $k_j$. 
In particular, the involving number at the hub $P_{h}$ is found 
numerically to scale as $P_h\sim N^2$, while the total involving 
number summed over all vertices is found numerically to scale as 
$\sum_{j=1}^N P_j \sim N^2 \log N$ (see Fig.10) \cite{correction}. 
So the normalized involving number at the hub $p_h \equiv P_h/\sum_j P_j$ 
behaves as $p_h \sim 1/\log N$, decaying much more slowly, compared 
with the normalized degree decreasing as $\sim N^{-0.5}$. 
This weak dependence on $N$ is comparable to the result that 
$v_{h,1} \sim {\cal O}(1)$ for large $N$. Therefore, the contribution 
of the hub to shortest paths is much larger than that 
of the naive estimate based on the normalized degree at the hub, 
${\cal O}(1/\sqrt{N})$. \\

In conclusion, we have considered the spectrum and eigenvectors 
of the adjacency matrix of the BA network, when bi-directional 
interaction is allowed. The density of the eigenvalues decays 
exponentially for small $|\lambda |$s, followed by power-law tails 
at both spectrum edges. This is different from the Wigner's formula 
appropriate to random graphs. We found that the largest two  
eigenvalues $\lambda_1$ and $|\lambda_N|$ depend on system size $N$ as 
$\sim N^{1/4}$ for large $N$, and the mass gap scales as $N^{-0.68}$. 
The eigenfunction corresponding to the largest eigenvalue $\lambda_1$ 
is strongly localized at the vertex with the largest degree, called the hub. 
The component of the normalized eigenfunction for $\lambda_1$ at the hub is 
independent of $N$, implying that the role of the hub in transport 
on the scale-free network becomes much important for larger 
system, and its contribution becomes much more dominant than expected 
according to the normalized degree at the hub, which scales as $N^{-1/2}$.
Therefore, it is very efficient in communication networks to construct 
central vertices, through which most of information traffic pass.\\ 

We would like to thank S.Y. Park for providing us with the Lanczos 
algorithm codes. This work is supported by grants No.2000-2-11200-002-3 
from the BRP program of the KOSEF. While writing this manuscript 
we learned of a work by Farkas $et$ $al.$ \cite{vicsek}, which overlaps 
some of our results.

\begin{figure}
\centerline{\epsfxsize=8.8cm \epsfbox{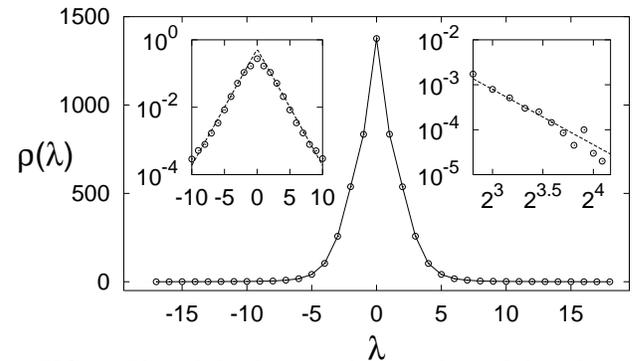}}
\caption{Plot of the density of eigenvalues of the adjacency matrix 
$\bf A$ versus eigenvalues for system size $N=5,000$ averaged over 40 
configurations. Left inset: Semi-logarithmic plot of the density 
versus eigenvalues to show the exponential decay for small $|\lambda |$s. 
Right inset: Double logarithmic plot of the density 
versus eigenvalues to show the power-law decay at spectrum 
edge. }
\label{fig1}
\end{figure}

\begin{figure}
\centerline{\epsfxsize=8.8cm \epsfbox{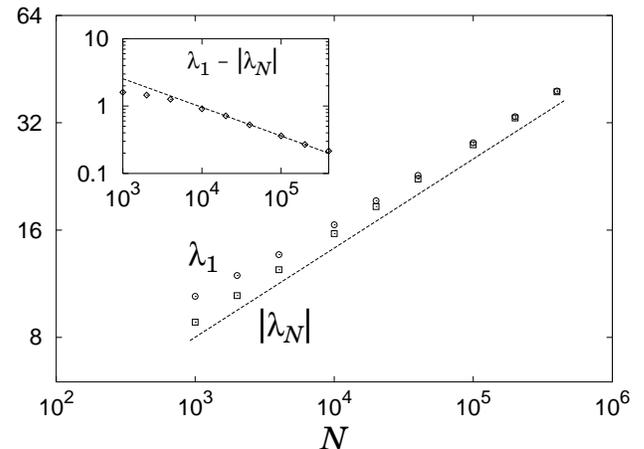}}
\caption{Double logarithmic plot of the first two largest eigenvalues 
$\lambda_1$ $(\circ)$ and $|\lambda_N|$ $(\Box)$ versus system size $N$. 
The dashed line 
has slope 0.25, drawn for the eye. Inset: Double logarithmic plot 
of the difference, $\lambda_1-|\lambda_N|$, versus system size $N$.}
\label{fig2}
\end{figure}

\begin{figure}
\centerline{\epsfxsize=8.8cm \epsfbox{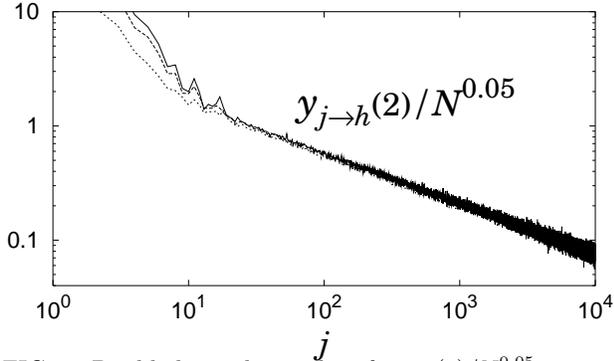}}
\caption{Double logarithmic plot of $y_{j\rightarrow h}(2)/N^{0.05}$ 
versus $j$ for different systems $N=1,000$, $5,000$ and $10,000$. 
The slope at the tail is estimated to be $-0.43$.} 
\label{fig3}
\end{figure}

\begin{figure}
\centerline{\epsfxsize=8.8cm \epsfbox{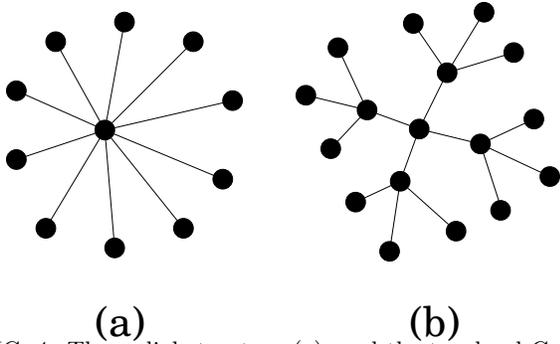}}
\caption{The radial structure (a), and the two-level Cayley tree 
structure with the coordination number $k=4$ (b).} 
\label{fig4}
\end{figure}

\begin{figure}
\centerline{\epsfxsize=8.8cm \epsfbox{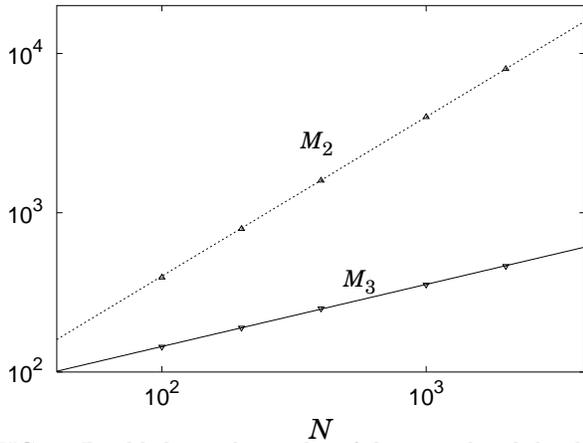}}
\caption{Double logarithmic plot of the second and the third moments 
${\cal M}_2$ and ${\cal M}_3$ versus system size $N$.
The dotted (solid) line has slope 1.0 (0.40), drawn for the eye.}  
\label{fig5}
\end{figure}

\begin{figure}
\centerline{\epsfxsize=8.8cm \epsfbox{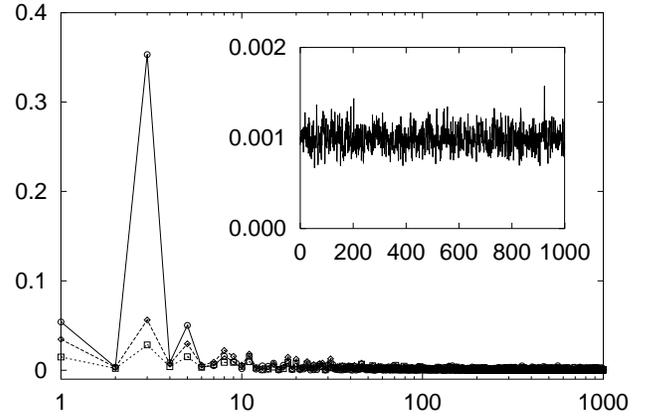}}
\caption{Plots of $v_{j,1}^2$ ($\circ$), the normalized degree 
$k_j/\sum_i k_i$ ($\Box$), and the normalized involving number 
$p_j$ ($\Diamond$) versus vertex index $j$ for $N=10^3$. 
Inset: The eigenfunction corresponding to the largest 
eigenvalue for the ER random network for $N=10^3$, showing that 
the eigenfunction is extended. } 
\label{fig6}
\end{figure}

\begin{figure}
\centerline{\epsfxsize=8.8cm \epsfbox{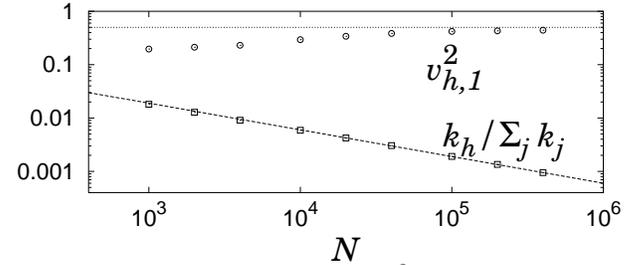}}
\caption{Double logarithmic plot of $v_{h,1}^2$ and the normalized 
degree $k_h/\sum_j k_j$ at the hub versus system size $N$. 
The dotted line $0.5$ is an asymptotic line of $v_{h,1}^2$.
The dashed line for the normalized degree has slope $-0.5$, drawn 
for the eye.} 
\label{fig7}
\end{figure}

\begin{figure}
\centerline{\epsfxsize=8.8cm \epsfbox{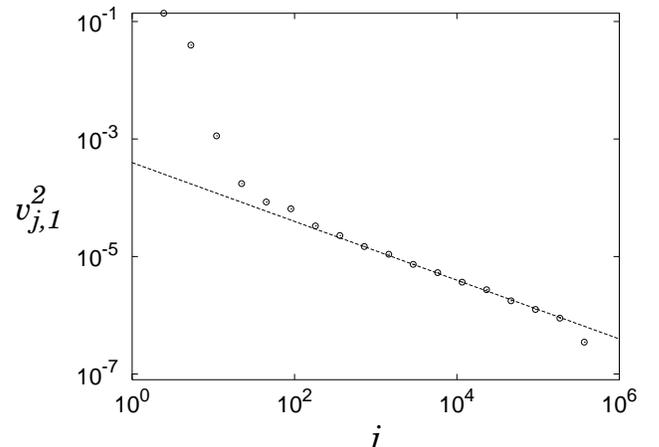}}
\caption{Double logarithmic plot of $v_{j,1}^2$ versus 
$j$ for $N=400,000$. The dashed line has slope $-0.5$, drawn 
for the eye.} 
\label{fig8}
\end{figure}

\begin{figure}
\centerline{\epsfxsize=8.8cm \epsfbox{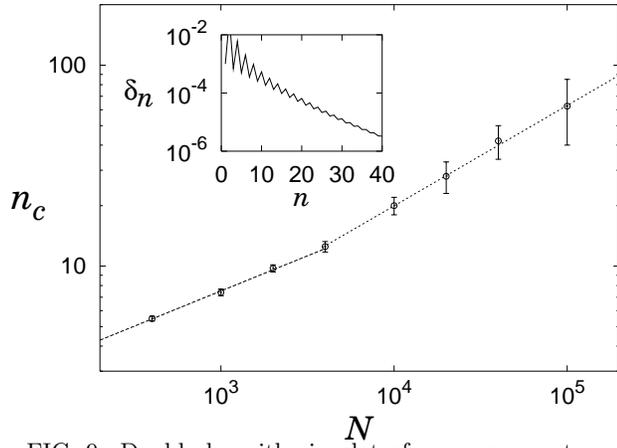}}
\caption{Double logarithmic plot of $n_c$ versus system size $N$. 
The dashed line has slope $0.35$ up to $N_c \approx 5,000$, 
and $0.50$ beyond $N_c$, drawn for the eye. 
Inset: Semi-logarithmic plot of $\delta_n$ versus number of steps $n$.} 
\label{fig9}
\end{figure}

\begin{figure}
\centerline{\epsfxsize=8.8cm \epsfbox{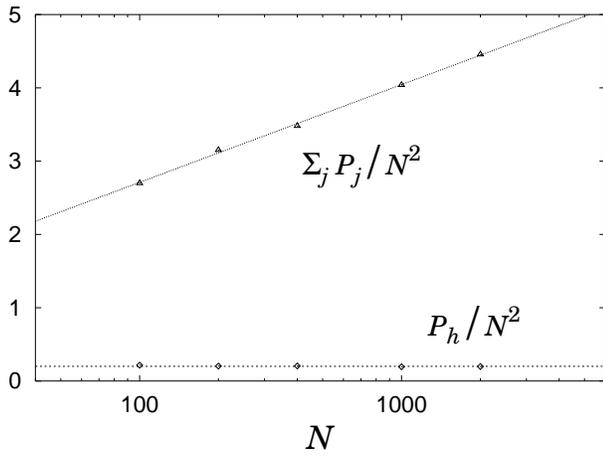}}
\caption{Semi-logarithmic plot of $P_h/N^2$ and $\sum_j P_j/N^2$.}
\label{fig10}
\end{figure}

\vfil\eject
\end{multicols}

\begin{thebibliography}{99}
\bibitem{correction} The numberical data may also be interpreted to follow a 
power-law $\sum_j P_j \sim N^{2.15}$.
\end{thebibliography}
\end{document}